\documentclass[%
 aip,
 amsmath, amssymb,
 reprint,%
]{revtex4-1}

\usepackage{graphicx}
\usepackage{dcolumn}
\usepackage{bm}

\usepackage[inkscapelatex=false]{svg}
\usepackage[utf8]{inputenc}
\usepackage[T1]{fontenc}
\usepackage{mathptmx}
\usepackage{etoolbox}

\makeatletter
\def\@email#1#2{%
 \endgroup
 \patchcmd{\titleblock@produce}
  {\frontmatter@RRAPformat}
  {\frontmatter@RRAPformat{\produce@RRAP{*#1\href{mailto:#2}{#2}}}\frontmatter@RRAPformat}
  {}{}
}%
\makeatother
\begin{document}

\preprint{AIP/123-QED}

\title{Ferroelectric AlBN Films by Molecular Beam Epitaxy}

\author{Chandrashekhar Savant*}
\email{cps259@cornell.edu}
\affiliation{\hbox{Department of Materials Science and Engineering, Cornell University, Ithaca, New York, 14853, USA}}

\author{Ved Gund}
\affiliation{\hbox{School of Electrical and Computer Engineering, Cornell University, Ithaca, New York, 14853, USA}}

\author{Kazuki Nomoto}
\affiliation{\hbox{School of Electrical and Computer Engineering, Cornell University, Ithaca, New York, 14853, USA}}

\author{Takuya Maeda}
\affiliation{\hbox{Kavli Institute at Cornell for Nanoscale Science, Cornell University, Ithaca, New York, 14853, USA}}

\author{Shubham Jadhav}
\affiliation{\hbox{School of Electrical and Computer Engineering, Cornell University, Ithaca, New York, 14853, USA}}

\author{Joongwon Lee}
\affiliation{\hbox{School of Electrical and Computer Engineering, Cornell University, Ithaca, New York, 14853, USA}}

\author{Madhav Ramesh}
\affiliation{\hbox{School of Electrical and Computer Engineering, Cornell University, Ithaca, New York, 14853, USA}}

\author{Eungkyun Kim}
\affiliation{\hbox{School of Electrical and Computer Engineering, Cornell University, Ithaca, New York, 14853, USA}}

\author{Thai-Son Nguyen}
 \affiliation{\hbox{Department of Materials Science and Engineering, Cornell University, Ithaca, New York, 14853, USA}}

\author{Yu-Hsin Chen}
 \affiliation{\hbox{Department of Materials Science and Engineering, Cornell University, Ithaca, New York, 14853, USA}}

\author{Joseph Casamento}
 \affiliation{\hbox{Department of Materials Science and Engineering, Cornell University, Ithaca, New York, 14853, USA}}

\author{Farhan Rana}
 \affiliation{\hbox{School of Electrical and Computer Engineering, Cornell University, Ithaca, New York, 14853, USA}}

\author{Amit Lal}
\affiliation{\hbox{School of Electrical and Computer Engineering, Cornell University, Ithaca, New York, 14853, USA}}

\author{Huili Grace Xing}
\affiliation{\hbox{Department of Materials Science and Engineering, Cornell University, Ithaca, New York, 14853, USA}}
\affiliation{\hbox{School of Electrical and Computer Engineering, Cornell University, Ithaca, New York, 14853, USA}}
\affiliation{\hbox{Kavli Institute at Cornell for Nanoscale Science, Cornell University, Ithaca, New York, 14853, USA}}

\author{Debdeep Jena}
\affiliation{\hbox{Department of Materials Science and Engineering, Cornell University, Ithaca, New York, 14853, USA}}
\affiliation{\hbox{School of Electrical and Computer Engineering, Cornell University, Ithaca, New York, 14853, USA}}
\affiliation{\hbox{Kavli Institute at Cornell for Nanoscale Science, Cornell University, Ithaca, New York, 14853, USA}}
\affiliation{\hbox{School of Applied and Engineering Physics, Cornell University, Ithaca, New York, 14853, USA}}

\begin{abstract}
We report the properties of molecular beam epitaxy deposited AlBN thin films on a recently developed epitaxial nitride metal electrode Nb$_2$N.  While a control AlN thin film exhibits standard capacitive behavior, distinct ferroelectric switching is observed in the AlBN films with increasing Boron mole fraction.  The measured remnant polarization $P_r \sim15$ \textmu C/cm$^{2}$ and coercive field $E_c \sim$ 1.45 MV/cm in these films are smaller than those recently reported on films deposited by sputtering, due to incomplete wake-up, limited by current leakage.  Because AlBN preserves the ultrawide energy bandgap of AlN compared to other nitride hi-K dielectrics and ferroelectrics, and it can be epitaxially integrated with GaN and AlN semiconductors, its development will enable several opportunities for unique electronic, photonic, and memory devices. 

\end{abstract}

\maketitle

\begin{quotation}

\end{quotation}


The piezoelectric and ultrawide direct energy bandgap properties of AlN make it the material of choice for radio frequency BAW filters and semiconductor photonic devices such as deep-ultraviolet LEDs and Lasers.\cite{jena2019new, 2023AlNPiezo, strite1993progress, bernardini1997spontaneous, muralt2009piezoelectric, kneissl2015III, wood2007polarization}  But it has not been possible to flip its spontaneous polarization with an external electric field. \cite{ye2021atomistic, PhysRevMaterials.5.044412} The current understanding is that the coercive field required to flip this polarization exceeds the critical breakdown field of this material\cite{ye2021atomistic, PhysRevMaterials.5.044412, fichtner2019alscn, wang2023dawn}, though the true critical breakdown field of AlN has not been measured to date.

In 2009, it was discovered that substituting group-III transition metal Sc atoms into Al sites of AlN dramatically boosted its piezoelectricity, leading to the rapid adoption of AlScN for BAW filters for cellphones.\cite{2023AlNPiezo}  In 2019, it was discovered that AlScN is ferroelectric,\cite{fichtner2019alscn} and in 2022 FerroHEMTs were realized by integrating AlScN epitaxially with GaN\cite{casamento2022ferrohemts, casamento2022transport, casamento2023review}.  The thermodynamic stable crystal structure of ScN of energy bandgap $\sim$1.2 eV is cubic rocksalt, and of AlN of energy bandgap $\sim$6.1 eV is hexagonal wurtzite.\cite{winiarski2020crystal, 2019ScNelectronicstructure, smith2001molecular, winiarski2020crystal, AlN2001prop}  Addition of Sc to AlN shrinks the energy bandgap and deforms the crystal structure from wurtzite towards cubic to an extent that it becomes possible for an external electric field to switch the polarization of AlScN, making it ferroelectric.\cite{fichtner2019alscn, wang2021piezoelectric, zhang2013tunable, noor2019ferroelectricity, liu2023doping, clima2021strain, ye2021atomistic}  This current understanding is under intense experimental and theoretical scrutiny.

In 2021, it was reported that introducing the non-transition metal group III Boron into AlN also led to ferroelectricity. \cite{PhysRevMaterials.5.044412}  While the AlN backbone stays the same and the dielectric constant is enhanced\cite{PhysRevMaterials.5.044412}, this route to ferroelectricity differs from the group-III transition metal ferroelectric AlScN \cite{calderon2023atomic, Dickey2023ISAF, fichtner2019alscn}(and its variants such as AlYN\cite{wang2023YAlN} or AlLaN\cite{rowberg2021structural, winiarski2020crystal}) in {\em three} essential ways: 1) the energy bandgap does not shrink substantially compared to AlN because the energy bandgap of BN variants (5.96 eV for h-BN, 6.4 eV for c-BN, 6.84 eV for wurtzite-BN) is similar to AlN\cite{kudrawiec2020bandgap, shen2017band, PhysRevMaterials.5.044412, suceava2023AlBNnonlinear, Savant2023AlBNMBE}, 2) the $d_{33}$ piezoelectric coefficient does not increase (it slightly {\em decreases} compared to AlN)\cite{suceava2023AlBNnonlinear, kudrawiec2020bandgap}, and 3) the chemical bonds do not involve d-orbitals - the competing crystal structure of BN is layered hexagonal (graphitic) with sp$^2$ bonds, and the less stable cubic and hexagonal wurtzite variants of BN have sp$^3$ bonds.\cite{kudrawiec2020bandgap}

Ferroelectric AlBN has only been observed to date in layers deposited by sputtering.\cite{PhysRevMaterials.5.044412, zhu2022wake, Fanhe2023ISAF, zhu2021strongly, Drury2023ISAF, calderon2023atomic, Dickey2023ISAF}  Do these properties carry over to layers deposited by molecular beam epitaxy (MBE)?  This work explores the growth and ferroelectric properties of AlBN by MBE with an eye towards its integration with the GaN and AlN semiconductor electronic and photonic ecosystem in the future.  The polarization properties of AlBN epitaxially grown on AlN or GaN are expected to be affected by the substrate, which (unlike Silicon) are themselves polar semiconductors. A metallic bottom electrode allows for the study of the AlBN properties decoupled from the substrate polarity.  

\begin{figure*}
\includegraphics[width=\textwidth]{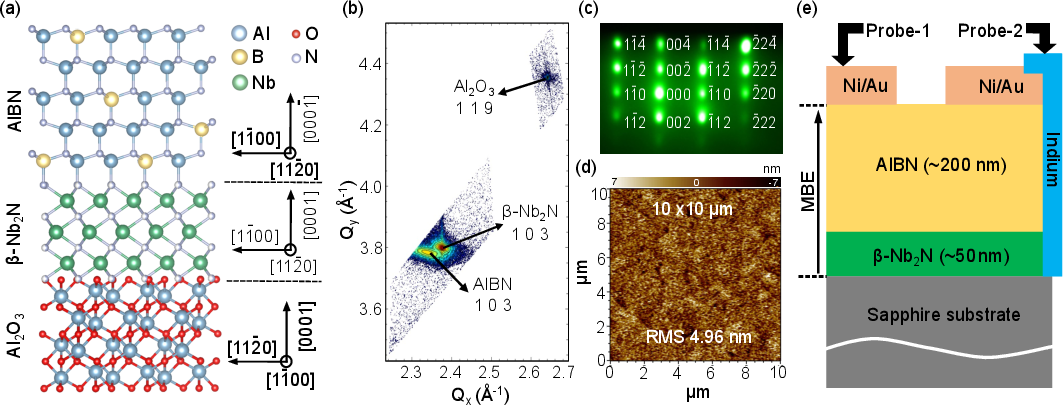}
\caption{\label{fig1} (a) Schematic of AlBN/$\beta$-Nb$_2$N/Al$_2$O$_3$ heterostructure showing epitaxial registry (b) XRD-RSM map showing AlBN 103, $\beta$-Nb$_2$N 103 and Al$_2$O$_3$ 119 Bragg reflections for sample with 18\% B (c) RHEED pattern of AlBN film with 18\% B along [11-20] zone axis, (d) AFM micrograph of AlBN film with 18\% B, and (e) cross-section of device used for electrical characterization.}
\end{figure*}

We summarize the methods before discussing the results.  For this study, the growths were performed in a Veeco® GenXplor nitrogen-plasma MBE system using 6N-purity Nb. B was evaporated from a Telemark® electron beam evaporation system, 6N-purity Al from a Knudsen effusion cell, and active N from a 6N5-purity N$_2$ gas at 1.95 sccm flow rate with a RF plasma power of 200 W.  In-situ monitoring of the growth surface was performed using a KSA Instruments reflection high energy electron diffraction (RHEED) apparatus with a Staib electron gun operating at 14.5 kV and 1.45 A. X-ray diffraction (XRD) reciprocal space mapping (RSM) was performed using a Panalytical Empyrean® diffractometer with Cu K$\alpha$1 radiation (1.54 \AA).  AFM imaging was performed using an Asylum Research Cypher ES system.  Analysis of chemical composition was performed using a Scienta-Omicron-ESCA-2SR X-ray photoelectron spectroscopy (XPS) instrument equipped with a 1486.6 eV Al K$\alpha$ source. A hemispherical analyzer was used to collect the photoelectrons to obtain high-resolution Al2p, B1s, N1s XPS spectra.  Ferroelectric P-E loops were obtained using Metal/AlBN/metal capacitors in a Sawyer-Tower setup by measuring the current response to positive-up-negative-down (PUND) voltage waveforms\cite{sawyer1930rochelle, traynor1997capacitor, gund2021temperature} from a Rigol DG1022 arbitrary waveform generator.  Piezo-response force microscopy (PFM) measurements were performed on the AlBN/$\beta$-Nb$_2$N/Al$_2$O$_3$ heterostructure sample in Asylum Research Cypher ES system using with a conductive tip at a 300 kHz frequency and V$_{AC}$ of 2 V. After PFM poling, the sample was etched in KOH to test the polarity of different poled regions.

Figure \ref{fig1}(a) schematically shows the layer structure grown by MBE for this study.  40-60 nm of Nb$_2$N was first deposited on a c-plane sapphire substrate by plasma-assisted MBE to serve as the epitaxial metallic bottom electrode.  The recently developed hexagonal Nb$_2$N of 300 K metallic resistivity $\rho \sim$ 44-57 $\mu$$\Omega$.cm is symmetry matched to wurtzite AlN, and is also nearly lattice matched to AlN, with only 1.4\% lattice mismatch.\cite{wright2023growth}.  Nb$_2$N, the most metal-rich phase of niobium nitride is stabilized and made phase pure in MBE by using a high growth temperature of 1150 $^{\circ}$C measured by a thermocouple.  Figure \ref{fig1}(b) shows the measured X-ray reciprocal space map 119 reflection of the sapphire substrate and the 103 reflection of Nb$_2$N.  Though Nb$_2$N is not lattice-matched to sapphire and is relaxed to serve as a substrate for AlBN, we note that it is also superconducting below $T \sim 2 $ K, making it of interest as a lattice-matched substrate for future AlN-based classical and quantum acoustodynamic devices.

Because N has a thermodynamic preference to bond to Al than to B\cite{hoke2007thermodynamic}, the epitaxial wurtzite phase AlBN films were grown on Nb$_2$N by supplying excess N compared to the net (Al + B) flux at a III/V flux ratio of $\sim$0.85, and the B mole fraction of AlBN was found to be controlled by the substrate temperature.\cite{Savant2023AlBNMBE} AlBN films of 0\% B (AlN control sample), and 4.7\% and 18\% B were deposited by MBE at growth substrate thermocouple temperatures of 600 $^{\circ}$C, 800 $^{\circ}$C, and 600 $^{\circ}$C respectively to obtain 200-250 nm thick epilayers directly after depositing the Nb$_2$N MBE layers as schematically shown in Figure \ref{fig1}(a).  Figure \ref{fig1}(b) shows that the measured X-ray 103 peak of the 18\% B content AlBN close to the 103 Nb$_2$N peak and far from the 119 Al$_2$O$_3$ peak, implying AlBN is of wurtzite phase, and it is in epitaxial registry with the Nb$_2$N. XRD-RSM maps of AlBN films with 0\%, and 4.7\% B (not shown) indicate similar crystal structure and an epitaxial registry of 30 degree orientation between hexagonal basal in-planes of AlBN, $\beta$-Nb$_2$N films and Al$_2$O$_3$.  

\begin{figure*}
\includegraphics[width=\textwidth]{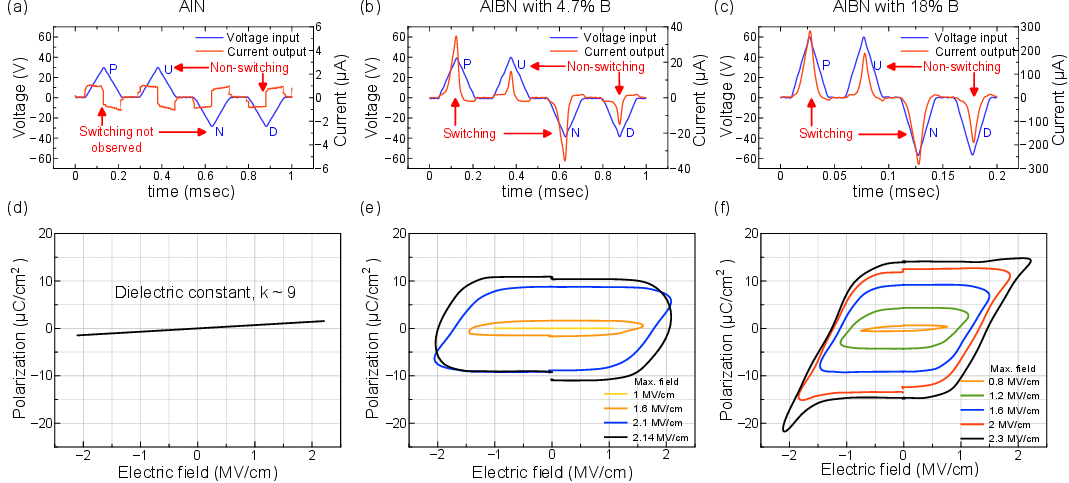}
\caption{\label{fig2} (a), (b), (c) triangular PUND waveform voltage pulses, output currents, and (d), (e), (f) P-E loops for AlBN with 0\%,  4.7\% and 18\% B samples, respectively.}
\end{figure*}

The RHEED image of AlBN films along [11-20] zone axis as shown in Figure \ref{fig1}(c) for AlBN with 18\% B confirms that AlBN films grow in the wurtzite structure.  The RHEED diffraction spots as labeled in \ref{fig1}(c) indicate Bragg reflections corresponding to first order 1 $\times$ 1 reconstruction for wurtzite structure. \cite{casamento2019ScN, henini2012MBEbook}  All Al$_{1-x}$B$_x$N films with x = 0\%, 4.7\%, 18\% show similar spotty wurtzite RHEED patterns, indicating a three-dimensional growth mode for the nitrogen-rich growth conditions used here for B incorporation in AlBN films.\cite{casamento2021ferroelectricity}  Figure \ref{fig1}(d) shows the AFM surface morphology; $\sim$150-250 nm AlBN films with 0\%, 4.7\% and 18\% B show similar morphologies with a rms roughness of $\sim$5 nm. 

Figure \ref{fig1}(e) shows a schematic of the device cross-section used for electrical characterization. The buried epitaxial $\beta$-Nb$_2$N film serves as a bottom-electrode. Top metal electrodes of Ni/Au were realized by ex-situ e-beam evaporation and lithographic patterning. Input waveforms were applied to circular electrodes of 40, 80, or 180 $\mu$m diameters (probe-1), and the resulting sense current was measured on a much larger electrode of size 300 $\mu$m $\times$ 8000 $\mu$m (probe-2) that was shorted to the bottom $\beta$-Nb$_2$N layer using indium for some pads, and by applying a large voltage pulse to ensure dielectric breakdown of the AlBN under the large electrode to others. Shorting of larger electrode (probe-2) with bottom $\beta$-Nb$_2$N was verified with a <8 $\Omega$ low dc resistance. A vertical capacitor thus forms between the smaller top-electrode and continuous bottom-electrode. 

The resulting capacitors connected in a Sawyer-Tower setup were subjected to positive-up-negative-down (PUND) waveforms with equal rise, fall, and wait times for continuous wave (CW) operation.  Figure \ref{fig2}(a) shows the current response in red of the AlN control capacitor to the PUND voltage waveform in blue.  This device exhibits only the displacement current $I = C dV/dt$ and no ferroelectricity.  The P-E curve obtained from the displacement current for this control sample in Figure \ref{fig2}(d) follows $P = \chi_{e} E$  where $\chi_{e}$ is the electric susceptibility, yielding a relative dielectric constant $k = 1 + \chi_{e} \sim 9$ similar to the reported low-frequency dielectric constant of AlN\cite{AlN2001prop, Jena2022quantum}.

The corresponding current responses of the Boron-containing AlBN capacitors are shown in Figures \ref{fig2}(b) and (c).  Riding on top of the displacement currents, additional (polarization switching + leakage) current components are observed in response to the P, N (switching) voltage waveforms, followed by (only leakage) currents in response to the U, D (non-switching) waveforms.  The polarization switching current density as a function of E-field is obtained by subtracting the leakage and displacement currents, from which the polarization charge $P$ is obtained by integrating the switching current.  The electric field $E$ is the applied voltage divided by AlBN layer thickness measured by cross-sectional SEM. The PUND testing was performed on AlBN samples of different B\%, under several measurement conditions (PUND frequencies, peak $E$-fields, pulsed waveforms) and on multiple devices to obtain the hysteretic $P-E$ loops similar to what is shown in Figures \ref{fig2}(e) and (f). 

The hysteresis observed in the $P-E$ loops for AlBN samples with 4.7\% and 18\% B are small at low electric fields, indicating incomplete switching. Increasing the electric field enlargens the loops; a completely switched ferroelectric with very few or no misoriented domains should exhibit a saturation of the remnant polarization $P_r$ for electric fields larger than the coercive field $E_c$. At higher electric fields, the remnant polarization and coercive fields converge to values of $P_r \sim10$ \textmu C/cm$^{2}$, $E_c \sim $1.71 MV/cm for 4.7\% B containing AlBN, and $P_r \sim 15$ \textmu C/cm$^{2}$ and $E_c \sim$1.45 MV/cm for 18\% B containing AlBN.  

\begin{figure}[t!]
\centerline{\includegraphics[width=1\columnwidth]{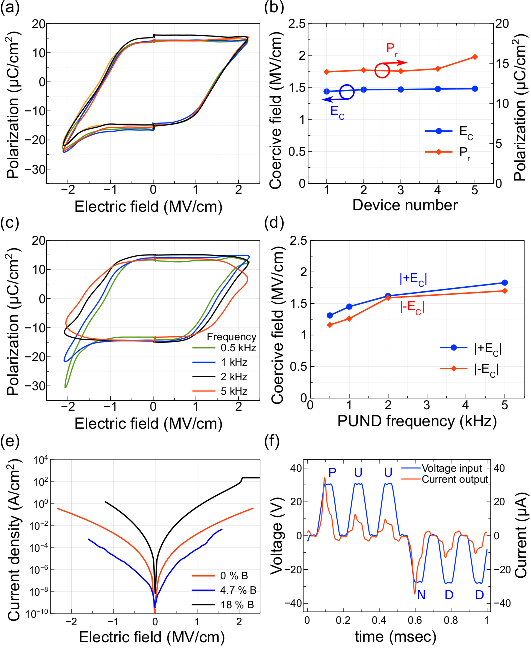}}
\caption{\label{fig3} (a), (b) P-E loops and $E_{c}$, $P_{r}$ values showing repeatability over multiple devices, (c), (d) P-E loops and $E_{c}$ values at various measurement frequencies on Al$_{0.82}$B$_{0.18}$N sample, (e) Current density, electric field plot for AlBN samples with 0\%, 4.7\%, 18\% B, and (f) a representative trapezoidal PUUNDD waveform voltage pulse, output current on Al$_{0.82}$B$_{0.18}$N sample.}
\end{figure}

Figure \ref{fig3}(a) shows the measured $P-E$ loops for various AlBN capacitors with 18\% B concentration, and Figure \ref{fig3}(b) shows the corresponding positive $E_c$ and $P_r$. Figure \ref{fig3}(c) shows the resulting loops for various frequencies of measurement, with the resulting positive and negative coercive fields shown in Figure \ref{fig3}(d) indicating a slight increase with frequency.  These measurements indicate a reasonable degree of uniformity and repeatability of the behavior.  The remnant polarization values saturate to $P_r \sim 15$ \textmu C/cm$^{2}$ as the frequency is increased.  Because the top electrode Ni/Au and the bottom electrode Nb$_2$N are not the same metal, an imprint difference in the positive and negative coercive fields is expected and observed in Figure \ref{fig3}(d) (and supplementary Figure S1), though a quantitative agreement would require more work in the future.   

Figure \ref{fig3}(e) shows the leakage current density in the AlBN capacitors as a function of the Boron concentration of AlBN.  The leakage currents of the control and 4.7\% Boron content AlBN stay at below mA/cm$^2$ levels at $E \sim 1$ MV/cm field, and that of the 18\% AlBN capacitor has a higher leakage.  Because with the increase of leakage current with electric field, it approaches the ferrolectric polarization switching current, polarization data for electric fields larger than 2 MV/cm are rejected for the devices measured here.  Figure \ref{fig3}(f) shows the polarization currents in the P and N pulses in response to trapezoidal voltage pulses.  Though there is leakage and displacement currents in the UU and DD pulses, the polarization switching currents in the P and N pulses are unambiguous.  Thus, we only retain data for electric fields for which such unambiguous polarization switching currents are observed. 

\begin{table}[ht]
\caption{\label{table1} Comparison of sputter-deposited and MBE-grown AlBN films with AlBN deposition method, bottom, top electrodes, B content, $E_{c}$, $P_{r}$.}
\begin{ruledtabular}
\begin{tabular}{cccccccc}
Deposition & Bottom/ & [B], & $E_{c}$,&$P_{r}$,&Comment\\
method &Top electrode &\% & MV/cm & $\mu$C/cm$^{2}$&\\
\hline
MBE & $\beta$-Nb$_2$N/ & 18\% & 1.45 & 15 & This work\\
 & Ni,Au & & Incomplete&wake-up & \\
MBE & $\beta$-Nb$_2$N/ & 4.7\% & 1.71 & 10 & This work\\
 & Ni,Au & &  Incomplete&wake-up & \\
Sputter & W/ &  7\% & 5.4 & 136 & Reference\cite{PhysRevMaterials.5.044412}\\
 & W & &  After&wake-up & \\
\end{tabular}
\end{ruledtabular}
\end{table}

Table \ref{table1} lists the ferroelectric properties of MBE AlBN films measured here in comparison to reported values for sputtered films \cite{PhysRevMaterials.5.044412, zhu2022wake} kindly shared by the authors of that work.  While the polarization switching of AlBN in the MBE films is unmistakable, the leakage currents and incomplete wake-up currently limit the maximum electric fields that can be applied for switching to a smaller coercive field and remnant polarization than the sputtered layers.\cite{zhu2022wake, Fanhe2023ISAF}  The lower coercive field and lower remnant polarization of the smaller loops are compatible with interfacing nitride ferroelectrics with conducting channels in GaN-based semiconductors because they remain below the typical mobile charge densities and breakdown fields of the semiconductor\cite{AlN2001prop, Jena2022quantum}. It must be pointed out that without further study with lower leakage currents and complete wake-up in AlBN layers, it is not possible to state with confidence the intrinsic $P-E$ loop of MBE-grown AlBN layers. 

\begin{figure*}
\includegraphics[width=\textwidth]{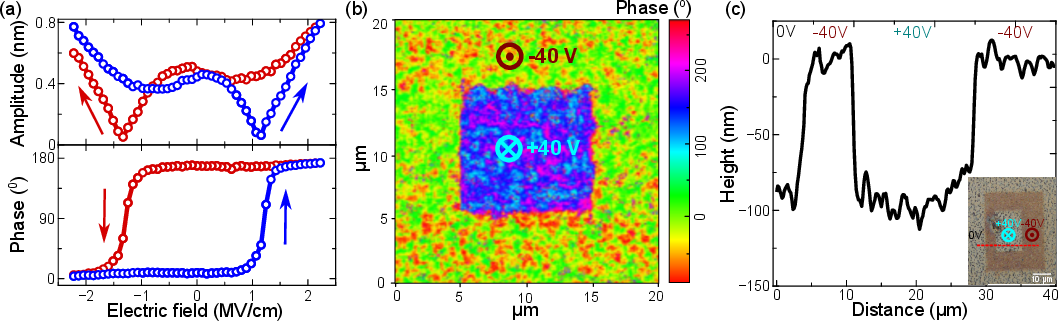}
\caption{\label{fig4} (a) Typical PFM amplitude and phase loops measured on AlBN/$\beta$-Nb$_2$N/Al$_2$O$_3$ sample with $\sim$ 18\% B measured at 300kHz and at room temperature. Butterfly-shaped amplitude loops and hysteretic phase loops with $\sim$180 degree phase shift are observed. (b) PFM phase contrast image pattern after writing at +/-40 V indicating electrically written domains of opposite polarizations. The inner 10 $\times$ 10 $\mu$m domain region written with +40 V bias on the tip has downward polarization, and the outer region written with -40 V bias has an upward polarization. (c) AFM line scan profile along the marked line in inset optical micrograph taken after KOH etching (45\% KOH, 25s) of AlBN sample with different PFM poled regions written at +/-40 V. The unpoled region without PFM writing and the innermost rectangle region written with +40 V bias have nitrogen polarity, as it etches more rapidly than the region written with -40V bias having metal polarity. 
}
\end{figure*}

To further probe the ferroelectric switching of the MBE AlBN films, PFM measurements were performed on the AlBN/$\beta$-Nb$_2$N/Al$_2$O$_3$ samples grown with B from effusion cell at $\sim$300 kHz, with the bottom $\beta$-Nb$_2$N electrode contacted with Ag paste, and the PFM conductive tip as the top electrode. Figure \ref{fig4}(a) shows the measured butterfly-shaped amplitude curve and the box-shaped hysteretic phase loop with $\sim$180 degree difference at zero fields, indicating that the polarity of the AlBN film is switched by the external electric field. The PFM coercive field value of $E_c \sim$1.22 MV/cm for 18\% B containing AlBN in Figure \ref{fig4}(a) is comparable to the PUND measured $E_c$ $\sim$1.45 MV/cm.  The amplitude and phase hysteresis loops were repeatable for several measurements at a fixed location, indicating repeatable switching between metal and nitrogen polarity.  Similar hysteretic loops were also observed at other locations of AlBN films grown by MBE. 

Figure \ref{fig4} (b) shows the PFM out-of-plane phase contrast image of domains after electrically writing the pattern at +/-40V bias on the PFM tip.  The inner 10 $\times$ 10 $\mu$m domain region written by +40V bias has downward polarization, and the outer region written by -40V has an upward polarization. The observed 180-degree phase difference across the clear domain boundary is characteristic of ferroelectrics. This result indicates that stable domains of opposite polarity can be written in the MBE-grown AlBN films. \cite{yun2022HZOPFM, wang2023dawn, wang2022epitaxial}  The region of the as-grown film without PFM poling is in-phase with the regions written by positive bias, whereas it is out-of-phase with the region written by negative bias [see supplementary Figures S3, S4(f), and S5(f) for details]. This indicates that applying a negative bias on the top electrode switches the polarity of the as-grown AlBN films. The writing of inverted patterns at another location on the sample and the stability of the polarity switched domains (detectable beyond 30 minutes of writing; see supplementary Figure S5) further demonstrate the ferroelectricity.

Finally, because the nitrogen-polar surfaces of wurtzite semiconductors such as AlN, GaN, and InN etch at higher rates than metal-polar surfaces in KOH or H$_3$PO$_4$\cite{zhuang2005wetreview, mariano1963ZnO, bickermann2007wet, fichtner2019alscn}, we subjected the PFM poled AlBN structures similar to that shown in Figure \ref{fig4}(b), (and supplementary Figure S4, S5) to wet etching in 45\% KOH and the surface profile was observed with AFM.  Figure \ref{fig4}(c) shows etched regions of $\sim$100 nm depth at the interface between regions of different polarity after KOH etching.  The AlBN film without PFM poling and the innermost rectangular region in inset Figure \ref{fig4}(c) written with +40 V bias on top PFM tip gets rapidly etched by the KOH, indicating its nitrogen polarity, whereas the regions written with -40 V bias on the top PFM tip etch slowly indicating its metal polarity. For reference, PFM and wet etching studies were also performed on the control 0\% B containing AlN/$\beta$-Nb$_2$N sample. Unlike AlBN film, the control AlN film did not show different etching rates at the regions biased with +/-50 V.


The modern theory of polarization (MTP) considers differential polarization between two structures and requires a known reference crystal structure to predict the spontaneous polarization of a given structure.\cite{resta1992theory, king1993theory, resta1994macroscopic} Using zinc-blende as a reference structure, Bernardini et al. determined the direction of the spontaneous polarization vector in the wurtzite nitrides to be along the [0 0 0 -1] crystallographic direction.\cite{PhysRevB.56.R10024, ambacher1999two}  In the last two decades this theory has explained various experimental phenomena relying on polarization discontinuity such as 2DEG and 2DHG in HEMT devices, internal fields, polarization doping, etc.

Recent calculations performed by Dreyer et al. using a layered hexagonal state as an initial reference state indicate different values and directions of spontaneous polarization in wurtzite nitrides, i.e., along the [0 0 0 1] crystallographic direction.\cite{dreyer2016correct}  Although both models can explain various experimental phenomena arising from polarization discontinuity, the actual direction and absolute polarization values in these ferroelectric wurtzite nitrides are under scrutiny.\cite{yassine2024modeling, wang2024rethinking}  The recent discovery of ferroelectricity in AlScN and AlBN has enabled the direct experimental determination of the spontaneous polarization in the wurtzite nitrides, wherein the experimentally observed values and direction are consistent with those predicted using layered hexagonal reference. \cite{yassine2024modeling, tang2023sub,  zhu2022wake, calderon2023atomic, dreyer2016correct}

The MBE AlBN film region written with a positive PFM tip bias, i.e., the electric field applied downward from top to bottom, gets etched more rapidly, suggesting its nitrogen polarity with a downward polarization direction along the [0 0 0 1] crystallographic direction. This is consistent with the model predicted using the layered hexagonal state as reference\cite{dreyer2016correct}.  However, a more detailed analysis of ferroelectric films' spontaneous, piezoelectric, and net polarization direction, film polarity, and strain state is required.  Direct experimental measurements using STEM, DPDC, and in-situ switching measurements of the ferroelectric wurtzite nitrides could help shed light on this rapidly emerging field.  Nonetheless, the above results (PUND, PFM, and wet etching) collectively provide strong evidence that polarization switching between states of opposite polarities occurs in the MBE-grown AlBN films by an external electric field and the AlBN films are ferroelectric.

It is encouraging to observe ferroelectric switching in MBE grown AlBN because of the intrinsic differences of such layers compared to AlScN and other potential nitride wurtzite ferroelectrics such as AlYN and AlLaN.\cite{calderon2023atomic, Dickey2023ISAF, fichtner2019alscn, wang2023YAlN, rowberg2021structural}  Like AlN, BN is also an ultrawide bandgap semiconductor in all crystalline forms, unlike the transition metal based ScN, YN, or LaN. \cite{kudrawiec2020bandgap, jena2019new, 2019ScNelectronicstructure, rowberg2021structural}  Therefore, AlBN is expected to remain ultrawide bandgap and maintain a large breakdown voltage compared to transition metal-based nitride ferroelectrics. \cite{kudrawiec2020bandgap, shen2017band, jena2019new, PhysRevMaterials.5.044412, suceava2023AlBNnonlinear, Savant2023AlBNMBE} Because B atom is lighter than Al, the thermal conductivity of AlBN, though lower than AlN due to alloy scattering of phonons, is expected to be higher than the heavier transition metal Sc, Y, or La alloys with AlN.  In addition, AlBN is expected to be more resistant to oxidation compared to AlScN and other transition metal-based nitride ferroelectrics since the transition metals possess an intrinsically higher affinity to form chemical bonding with oxygen in preference to nitrogen.\cite{casamento2020oxygen}  These differences indicate that AlBN has the potential to be an electronically, thermally, and chemically robust ferroelectric suited for harsh environments and high-temperature operation.

\vspace{5mm}
\section*{Supplementary Material}
See the supplementary material for imprint in the coercive field, leakage current density-electric field plots over multiple devices, as well as additional PFM measurements, poling, and KOH etching test results for film polarity check illustrating ferroelectric switching in the MBE AlBN films.
\vspace{-2mm}

\vspace{2mm}
\begin{acknowledgments}
\vspace{-2mm}
This work was supported as part of the Ultra Materials for a Resilient Energy Grid (epitaxy and device fabrication), an Energy Frontier Research Center funded by the U.S. Department of Energy, Office of Science, Basic Energy Sciences under Award $\#$DE-SC0021230 and in part by ARO Grant $\#$W911NF2220177 (characterization). The authors acknowledge the use of the Cornell NanoScale Facility (CNF) a member of the National Nanotechnology Coordinated Infrastructure (NNCI), which is supported by the National Science Foundation (NSF Grant NNCI-2025233). This work made use of the Cornell Center for Materials Research (CCMR) Shared Facilities which are supported through the NSF MRSEC program (DMR-1719875). The authors would like to thank Jimy Encomendero and Hardik Murarka from Cornell University for insightful discussions. The authors would also like to acknowledge John Hayden and Professor Jon-Paul Maria from The Pennsylvania State University for providing sputtered AlBN data.
\end{acknowledgments}
\vspace{-2mm}

\section*{Author Declarations}
\subsection*{Conflict of Interest}
The authors have no conflicts to disclose.

\section*{Data Availability}
The data that support the findings of this study are available from the corresponding author upon reasonable request.

\section*{References}
\bibliography{Ferroelectricity_in_MBE_AlBN_clean}

\end{document}